\documentclass[groupedaddress,superscriptaddress,aps,showpacs,amsmath,amssymb,floatfix, prl,twocolumn,a4]{revtex4-1}

 \usepackage{graphicx,float}
 \usepackage{subfigure}
\usepackage{bm}
\usepackage{mathrsfs}
\usepackage{txfonts}
\usepackage{CJK}
\usepackage[amssymb]{SIunits}
\usepackage{epsfig}
\usepackage{lipsum}
\usepackage{color}
\usepackage{textcomp}
\usepackage{placeins}

 \newenvironment{SChinese}{%
  \CJKfamily{gbsn}%
 \CJKtilde
  \CJKnospace}{}

\allowdisplaybreaks[2]

\begin{document}

\begin{CJK}{UTF8}{} 
\begin{SChinese}

\title{Squeezing giant spin states via geometric phase control in cavity-assisted Raman transitions}

 \author{Keyu Xia (夏可宇)}  %
 \email{keyu.xia@mq.edu.au}
 \affiliation{ARC Centre for Engineered Quantum Systems, Department of Physics and Astronomy, Macquarie University, NSW 2109, Australia}
 \affiliation{College of Engineering and Applied Sciences, Nanjing University, Nanjing 210008, China}


\date{\today}

\begin{abstract}
  Squeezing ensemble of spins provides a way to surpass the standard quantum limit (SQL) in quantum metrology and test the fundamental physics as well, and therefore attracts broad interest. Here we propose an experimentally accessible protocol to squeeze a giant ensemble of spins via the geometric phase control. Using the cavity-assisted Raman transitions in a double $\Lambda$-type system, we realize an effective Dicke model. Under the condition of vanishing effective spin transition frequency, we find a particular evolution time where the cavity decouples from the spins and the spin ensemble is squeezed considerably. Our scheme has the potential to improve the sensitivity in quantum metrology with spins by about two orders.
\end{abstract}

\pacs{42.50.Lc, 42.50.Dv, 42.50.Pq, 42.50.Nn }


\maketitle

\end{SChinese}
 \end{CJK}
 
 Spins, due to the merit of their long decoherence, have been widely used for ultrasensitive sensing of various signals \cite{NatPhys.3.227,NatPhys.10.21,Science.339.561,PhysRevX.4.021045,PhysRevLett.110.160802,PhysRevLett.112.160802,PhysRevLett.110.130802,PhysRevX.5.041001,NatCommun.6.8251,PhysRevA.92.043409}. However, the precision of the conventional measurement with spins is bounded by the shot noise or the SQL \cite{Science.306.1330,Science.344.1486}. Quantum spin squeezing and entanglement can surpass the SQL and therefore boost the sensitivity in quantum measurements to approach the Heisenberg limit \cite{Science.306.1330,PhysRep.509.89}. 
 
 To exploit the power of the spin-squeezed state (SSS), various methods have been proposed using quantum measurement \cite{AtomicSpinSqu2,AtomicSpinSqu1,AtomicSpinSqu4}, quantum bath engineering \cite{AtomicSpinSqu3}, converting entanglement to squeezing \cite{PhysRevLett.109.173603} and cavity feedback \cite{CavityFeedback1,CavityFeedback2}, typically for atomic ensembles. The state-of-the-art experiment has achieved $20~\deci\bel$ squeezing of half a million ultracold Rb atoms in a natural trap \cite{AtomicSpinSqu2}. Recently, Bennett et al. show the potential to squeeze $100$ nitrogen-vacancy (NV) spins in diamond via the Tavis-Cummings interaction with a nanomechanical resonator, mediated by strain \cite{PhysRevLett.110.156402}. Their scheme inevitably and sensitively suffers to the large thermal excitation of mechanical resonator. Zhang's and our works show that the NV centers can also couple to a mechanical resonator mediated by a giant magnetic gradient and the geometric phase control can be used to squeeze NV centers. Taking the merit of the geometric phase protocol robust again various noises, the squeezing is immune to thermal excitation \cite{PhysRevA.92.013825,KXJT}. However, the giant magnetic gradient causes large Zemman splitting in NV centers and is highly localized in nanometer region. As a result, the available number of spins is limited up to $20$ \cite{PhysRevA.92.013825,KXJT}. Cavity-assisted Raman transition (CART) has been proposed and then demonstrated for Dicke model quantum phase transitions \cite{PhysRevA.75.013804,PhysRevLett.113.020408,SuperfluidGas}. Here we aim to provide an experimentally feasible scheme to squeeze millions or even trillions spins using CART.
 
 In this letter, we propose a scheme for squeezing in a transient way a large ensemble of spins in an optical cavity via the geometric phase control, avoiding the complex configuration in squeezing spins via quantum measurement, quantum bath engineering or feedback. We couple the ensemble of ultracold alkali atoms or negatively charged silicon-vacancy (SiV$^-$) color centers in diamond or a superfluid gas formed in Bose-Einstein condensate (BEC) to the cavity. Using CART, we create an effective Dicke model for the spin-photon interaction. In a special arrangement, the effective resonance frequency, $\omega_c$, of the cavity is much larger than the effective transition frequency of the spins. At a particular time, $t=2\pi/\omega_c$, the spin and cavity decouples. At the same time, the ensemble of spins accumulates a geometric phase due to the collective interaction with the cavity and are collectively twisted along one axis of the Bloch sphere of spins. As a result, the cavity squeezes the spins considerably. Because the spins can be optically initialized to their ground state and the thermal excitation of the optical cavity is vanishing small even at room temperature, our scheme has an advantage that the thermal noise can be neglected in squeezing.  
  
 \begin{figure}
  \includegraphics[width=0.6\linewidth]{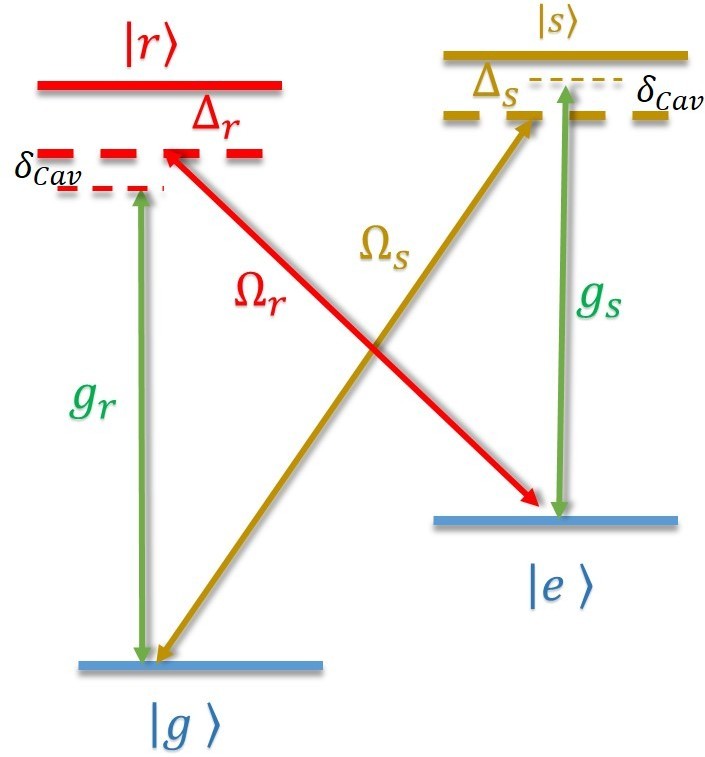} \\
  \caption{(Color online) Level diagram for showing two CARTs. We consider a cavity QED system in which an ensemble of spins (cold atoms or SiV centers or BEC) with double $\Lambda$ configuration is trapped in a good cavity. In combination with the cavity mode, two classical laser fields, $\Omega_r$ and $\Omega_s$, (red and brown) drives the spins to form Raman transitions between states $|e\rangle$ and $|g\rangle$.} \label{fig:Level}
 \end{figure}
 
 We start the discussion of our work by describing the system. Our configuration is a cavity electrodynamics (QED) system in which an ensemble of $N_a$ double $\Lambda$-type systems is trapped. The level diagram of the system is depicted in Fig. \ref{fig:Level}. Each $\Lambda$-type system has two optical excited states $|r\rangle$ and $|s\rangle$, and two metastable states $|g\rangle$ and $|e\rangle$.
 The state  $|j\rangle$ has energy $\hbar \omega_j$ ($j=r,s,g,e$). We assume that the excited states, $|r\rangle$ ($|s\rangle$) decay to the two ground states, $|g\rangle$ and $|e\rangle$, with the rates of $\gamma_{rg}$ and $\gamma_{re}$ ($\gamma_{sg}$ and $\gamma_{se}$), respectively.
 The cavity mode, $\hat{c}$, with resonance frequency $\omega_\text{cav}$ and decay rate $\kappa$, drives the transition $|g\rangle \leftrightarrow |r\rangle$ ($|e\rangle \leftrightarrow |s\rangle$) with strength $g_r$ ($g_s$). The classical laser fields drive atomic transitions $|e\rangle \leftrightarrow |r\rangle$ and $|g\rangle \leftrightarrow |s\rangle$ with Rabi frequency $\Omega_r$ and $\Omega_s$, respectively and detuning $\Delta_r=(\omega_r-\omega_e)-\omega_{lr}$ and $\Delta_s=(\omega_s-\omega_g)-\omega_{ls}$, respectively. $\omega_{lr}$ and $\omega_{ls}$ are the carrier frequencies of the laser fields $\Omega_r$ and $\Omega_s$. The paired interaction, $g_r$ and $\Omega_r$, $g_s$ and $\Omega_s$, forms two CARTs. Each CART drives the transition between two ground states. Combining these two CARTs, we obtain the Dicke Hamiltonian \cite{PhysRevA.75.013804} which is the key of our geometric phase control. 
  
 Before we go to the model, lets first briefly discuss three possible implementations using ultracold alkali atoms, SiV$^-$ centers in diamond or a superfluid gas. All three systems for implementations can be effectively treated as an ensemble of spin-$1/2$ systems in the Dicke model.
 As an example, we consider an ensemble of ultracold $^{87}\text{Rb}$ atoms for the first implementation \cite{PhysRevLett.113.020408,AtomData}. We choose $|r\rangle=|5^2P_{3/2},F^\prime=2,m_{F^\prime}=1\rangle$, $|s\rangle=|5^2P_{3/2},F^\prime=2,m_{F^\prime}=2\rangle$, $|g\rangle=|5^2S_{1/2},F=1,m_{F}=1\rangle$ and $|e\rangle=|5^2S_{1/2},F=2,m_{F}=2\rangle$ in the $D_2$ line of $^{87}\text{Rb}$ atom. According to atomic data \cite{AtomData}, the dipole moments are $d_{rg}=d_{re}=-\sqrt{1/8} d$ for the transitions $|r\rangle\leftrightarrow |g\rangle$ and $|r\rangle\leftrightarrow |e\rangle$, $d_{sg}=\sqrt{1/4} d$ for $|s\rangle\leftrightarrow |g\rangle$, and $d_{se}= \sqrt{1/6} d$ for $|s\rangle\leftrightarrow |e\rangle$, with $d=3.584\times 10^{-29} ~\coulomb\cdot \meter$. In such configuration, the cavity mode can be a linear-polarized field and the cavity-atom interaction is strong due to the large dipole-dipole moments. Other hyperfine levels can be effectively decoupled due to the large detuning which can also be adjusted with a constant magnetic field $B_c$ \cite{PhysRevLett.113.020408}.  The each excited state decays at a rate of $\gamma \sim 2\pi \times 6~\mega\hertz$ \cite{PhysRevLett.113.020408,AtomData}, yielding $\gamma_{rg}=\gamma_{re}=2\pi \times 3~\mega\hertz$, $\gamma_{rg}=2\pi \times 3.6~\mega\hertz$, and $\gamma_{se}=2\pi \times 2.4~\mega\hertz$ for different branches. 
 Interestingly, we can also squeeze an ensemble of solid-state spins, SiV$^-$ centers in diamond trapped in a cavity \cite{PhysRevLett.112.160802}. The SiV$^-$ centers in diamond cut with $\{111\}$ surface have shown a double $\Lambda$-type configuration \cite{PhysRevLett.113.263601,PhysRevLett.113.263602,PhysRevLett.112.036405}. To use SiV centers for our scheme, we take $|s\rangle=|^2\text{\bf E}_u,e_-^u,\uparrow\rangle$, $|r\rangle=|^2\text{\bf E}_u,e_-^u,\downarrow\rangle$, $|e\rangle=|^2\text{\bf E}_g,e_+^u,\uparrow\rangle$, $|g\rangle=|^2\text{\bf E}_g,e_+^u,\downarrow\rangle$, respectively \cite{NJP.17.043011}. The relaxation rate, $\Gamma$, of the spin ground state is negligible ($2.4 ~\milli\second$), but the pure dephasing, $\Gamma_\phi$, is about $2\pi \times 3.5~\mega\hertz$ \cite{PhysRevLett.113.263601,PhysRevLett.113.263602}. While, the relaxation of the optical excited states, $|r\rangle$ and $|s\rangle$, is negligible at cryogenic temperature \cite{NJP.17.043011}. We assume $d_{rg}=d_{re}=d_{sg}=d_{se}$. At $T=1~\kelvin$, we can take $\gamma_{rg}=\gamma_{re}=\gamma_{sg}=\gamma_{se}=2\pi \times 3.7~\mega\hertz$.
 More remarkably, our protocol can squeeze the momentum of a superfluid gas which can also construct the double $\Lambda$-type configuration \cite{SuperfluidGas}, taking $|r\rangle = |\pm \hbar k, 0\rangle^\prime$, $|s\rangle = |0,\pm \hbar k\rangle^\prime$, $|g\rangle=|0,0\rangle$ and $|e\rangle= |\pm \hbar k, \pm \hbar k\rangle$. The Dicke model driving the effective transition between $|0,0\rangle$, the atomic zero-momentum state, and $|\pm \hbar k, \pm \hbar k\rangle$, the symmetric superposition of momentum states can be created via the CART. The effective energy of the cavity and the spin can be controlled via the optical trapping potential, the photon-spin coupling, the detuning $\Delta_c$ and the atom-induced dispersive shift of the cavity resonance $UB$ \cite{SuperfluidGas}. The energy of the state  $|\pm \hbar k, \pm \hbar k\rangle$ is lifted relative to the state $|0,0\rangle$ by twice the recoil energy that $\omega_q=2\pi\times 28.6~\kilo\hertz$ \cite{SuperfluidGas}. While the effective energy, $\hbar\omega_c=\hbar \Delta_c - UB$ is typically much larger than $\hbar\omega_q$. In the experiment, the single-atom coupling $\eta>2\pi \times 0.9~\kilo\hertz$ is achieved. In the end of our numerical investigation, we will numerically evaluate the squeezing parameter as a function of the number of spins and then estimate the achievable squeezing degree for a large ensemble by fitting the numerical data.
 
 We now go to derive the Dicke Hamiltonian governing the evolution of system.  We  transform the system into the interaction picture by introducing the unitary transformation $\hat{U}(t)=\exp(-iH_0 t)$ with $H_0=\sum_j \omega_g |g_j\rangle \langle g_j| +  \omega_e |e_j\rangle \langle e_j| + (\omega_{lr}+ \omega_e) |r_j\rangle \langle r_j| + (\omega_{ls}+ \omega_g) |s_j\rangle \langle s_j| + \omega_\text{cav}^\prime \hat{c}^\dag \hat{c}$, as in \cite{PhysRevA.75.013804}. We set $\omega_{ls}-\omega_{lr}=2(\omega_e- \omega_g)$ that $\omega_\text{cav}^\prime=\omega_{lr}+(\omega_e-\omega_g)= \omega_{ls}-(\omega_e-\omega_g)$.
Thus we obtain the Hamiltonian in the interaction picture,
\begin{equation}
 \begin{split}
  H = \delta_\text{cav} & \sum_j (\Delta_r |r_j\rangle \langle r_j| + \Delta_s |s_j\rangle \langle s_j|) \; \\ 
  & + \sum_j \left(g_r e^{-ik r_j}\hat{c}^\dag |g_j\rangle \langle r_j| + g_s e^{-ik r_j}\hat{c}^\dag |e_j\rangle \langle s_j| + H.c. \right) \;\\ 
  & + \sum_j \left(\frac{\Omega_r}{2} e^{ik_{lr} r_j} |r_j\rangle \langle e_j| + \frac{\Omega_s}{2}  e^{ik_{ls} r_j} |s_j\rangle \langle g_j| + H.c.\right) \; ,
 \end{split}
\end{equation}
 where $k=\omega_\text{cav}/C$, $k_{lr}=\omega_{lr}/C$ and $k_{ls}=\omega_{ls}/C$ with $C$ is the light velocity in vacuum are the wave vector of the cavity mode and the classical laser fields, $r_j$ is the position of the $j$th spin. We assume $k\approx k_{lr} \approx k_{ls}$. Taking $|\Delta_{r,s}| \gg \Omega_{r,s}, g_{r,s}, \gamma$, we adiabatically eliminate the optical excited states $|r_j\rangle$ and $|s_j\rangle$, and neglect the constant energy terms to arrive at the Dicke model Hamiltonian for the collective coupling of the ground states $|g_j\rangle$ and $|e_j\rangle$ \cite{PhysRevA.75.013804,SupplementaryInf}, 
 \begin{equation} \label{eq:HDicke}
   H_\text{Dicke} = \omega_c \hat{c}^\dag \hat{c} + \omega_q J_z + 2\sqrt{N_a}\lambda (\hat{c}^\dag+ \hat{c})\bar{J}_x \;,
 \end{equation}
 where $\omega_c= \delta_\text{cav} -\frac{1}{2}N_a \left(\frac{|g_r|^2}{\Delta_r} + \frac{|g_s|^2}{\Delta_s} \right)$, $\omega_q = \frac{|\Omega_s|^2}{4\Delta_s}-\frac{|\Omega_r|^2}{4\Delta_r}$ caused by the ac Stark shifts. Namely, the two-photon detuning in the CARTs is $\delta_\text{cav}$. In Hamiltonian Eq. (\ref{eq:HDicke}), we define the collective operators for the spins, $J_z=\sum_j(|e_j\rangle \langle e_j|-|g_j\rangle \langle g_j|)/2$, $J_+=J_-^\dag=\sum_j |e_j\rangle \langle g_j|$ and $\bar{J}_x=(J_+ + J_-)/2\sqrt{N_a}$. 
 Here we, for our purpose of squeezing spins, choose $\frac{|g_r|^2}{\Delta_r}=\frac{|g_s|^2}{\Delta_s}$ and $\lambda=\frac{\Omega_r^* g_r}{2\Delta_s}=\frac{\Omega_s g_s^*}{2\Delta_s}$ by controlling the detuning and the classical driving. Essentially, these conditions requires $\Delta_r/\Delta_s= |d_{rg}|^2/|d_{se}|^2$ and $\Omega_r/\Omega_s=d_{rg}/d_{se}$ when the dipole moments $d_{rg,se}$, $g_{r,s}$ and $\Omega_{r,s}$ are real numbers. As a results, $\omega_q=0$ is obtained. We will also investigate the case of $\omega_q \neq 0$ for a general discussion of squeezing BEC.
 We can consider the ensemble of spins as a resonator with annihilation operator $\hat{a}$ under the Holstein-Primakoff (HP) transformation that $J_z=(\hat{a}^\dag \hat{a} - \mathscr{N}/2)$, $J_+=\hat{a}^\dag \sqrt{\mathscr{N}-\hat{a}^\dag \hat{a}}$, $J_-= \sqrt{\mathscr{N}-\hat{a}^\dag \hat{a}}\hat{a}$, and $\bar{J}_x=(\hat{a}^\dag \sqrt{\mathbf{I}-\hat{a}^\dag \hat{a}/N_a} + \sqrt{\mathbf{I}-\hat{a}^\dag \hat{a}/N_a}\hat{a})/2$ \cite{HPTransf,DickeQPT1}, where $\mathscr{N}=N_a \mathbf{I}$. In the ideal case of $\omega_q=0$, we rewrite the Hamiltonian in the interaction picture of $\omega_c \hat{c}^\dag\hat{c}$ as 
\begin{equation} \label{eq:V}
 V_x=2\sqrt{N_a}\lambda( e^{i\omega_c t}\hat{c}^\dag + e^{-i\omega_c t}\hat{c}) \bar{J}_x \;.
\end{equation}

Now we go to the geometric phase control of the evolution of the system. By applying the Magnus's formula \cite{MagnusFormula}, the dynamics for the system is governed exactly, in the absence of decoherence, by the unitary operator $U_x(t)=e^{iN_a\theta(t) \bar{J}_x^2} e^{2\lambda/\omega_c(\alpha(t)\hat{c}^\dag -\alpha^*(t)\hat{c})\bar{J}_x}$, where $\alpha(t)=1-e^{i\omega_c t}$, and $\theta(t)=\left(\frac{2\lambda}{\omega_c}\right)^2 (\omega_c t -\sin\omega_c t)$. $\theta(t)$ is the accumulated geometric phase only dependent on the global geometric features of operators and is robust against random operation errors \cite{PhysRevLett.90.160402}. Note that the spin-cavity coupling is modulated quickly by the periodic function $\alpha(t)$. At $t_m=2m\pi/\omega_c$ for an integer $m$, $\alpha(t_m)$ vanishes, $\theta(t_m)=2m\pi \left(\frac{2\lambda}{\omega_c}\right)^2$ and the spins decouple from the cavity. As a result, the evolution operator for the spin ensemble takes an explicit form,
\begin{equation} \label{eq:Utm}
 U_x(t_m)=e^{iN_a\theta(t_m) \bar{J}_x^2} \;.
\end{equation}
Given the initial state $|\Psi(0)\rangle$ for the spin ensemble, the generated state after one period, i.e. at $t_1$ is $|\Psi(t_1)\rangle=U_x(t_1)|\Psi(0)\rangle$. It is noticeable that the squeezing degree of the SSS only depends on the accumulated geometric phase $\theta(t_1)$, which can be adjusted with the classical driving and the detuning. 

The power of our protocol in squeezing spins is limited by the discrepancy of $\omega_q$ from zero and the decoherence of system.
Although we set $\omega_q=0$ for the analysis of ideal geometric phase control, the protocol actually works efficiently when $\omega_c \gg \omega_q$. In comparison with the protocol using a mechanical resonator to enable the geometric phase control \cite{PhysRevA.92.013825,KXJT}, the crucially detrimental thermal noise is negligible in our scheme because the thermal excitation of the optical cavity is vanishing small and the spins can be optically polarized in the ground state $|g_j\rangle$. The decay of excited states $|r_j\rangle$ and $|s_j\rangle$ can introduce some coherence to the evolution via CARTs but is suppressed by the large detuning \cite{RamanModel}. Threfore, the decay of the cavity is the main decoherence source. Another decoherence source is the pure dephasing, $\Gamma_\phi$, of the ground state $|e_j\rangle$. To taking into account the influence of the imperfection in $\omega_q$ and the decoherence, we numerically solve the quantum Langevin equation in the HP picture \cite{RamanModel,SupplementaryInf},
\begin{equation} \label{eq:MEq}
 \begin{split}
 \partial \rho/\partial t = & -i [H_\text{Dicke},\rho]   + \mathscr{L}(\sqrt{\Gamma_\phi/2}J_z)\rho + \mathscr{L}_c(\sqrt{\kappa}\hat{c})\rho \;,
 \end{split}
\end{equation}
where $\mathscr{L}_c(\hat{A})\rho =\hat{A}\rho \hat{A}^\dag - \frac{1}{2} \hat{A}^\dag \hat{A} \rho -\frac{1}{2} \rho \hat{A}^\dag \hat{A}$.

In our three implementations, the dark states of spins are rarely excited, thanks to the small inhomogeneous broadening of the excited state. Therefore, we focus on the symmetric states with the total spin $J=N_a/2$. The state of spin ensemble can be fully described by set of the Dicke state $|J,m\rangle$ with $m\in \{-J,-J+1,\cdots, J-1,J\}$ in the spin picture, which is equivalent to the Fock state $|J+m\rangle$ in the Bosonic or HP picture. In the later, the squeezing degree of spin states $\{|g\rangle, |e\rangle\}$ of spin ensemble can be evaluated by the squeezing parameter defined by Wineland et al. as $\xi_R^2=\left(\frac{N_a}{2|\langle \vec{J}\rangle|} \right)^2 \xi_s^2$ \cite{PhysRep.509.89}, where $|\langle \vec{J}\rangle|=\sqrt{\langle J_x\rangle^2 + \langle J_y\rangle^2 + \langle J_z\rangle^2}$ and the squeezing parameter $\xi_s^2=1+2\langle \hat{a}^\dag \hat{a}\rangle -2 \frac{\langle (\hat{a}^\dag \hat{a})^2\rangle}{N_a}-2|\langle \bar{J}_x^2\rangle|$ is given by Kitagawa and Ueda \cite{PhysRep.509.89}. The squeezing is optimal at $\theta_\text{opt}=6^{-1/6}(N/2)^{-2/3}$ \cite{KXJT}. %
Correspondingly, the phase uncertainty in quantum metrology with such SSS can be reduced down to $\delta\phi=\xi_R/\sqrt{N}$, improved by a factor of $\xi_R$. 

\begin{figure}
  \includegraphics[width=1\linewidth]{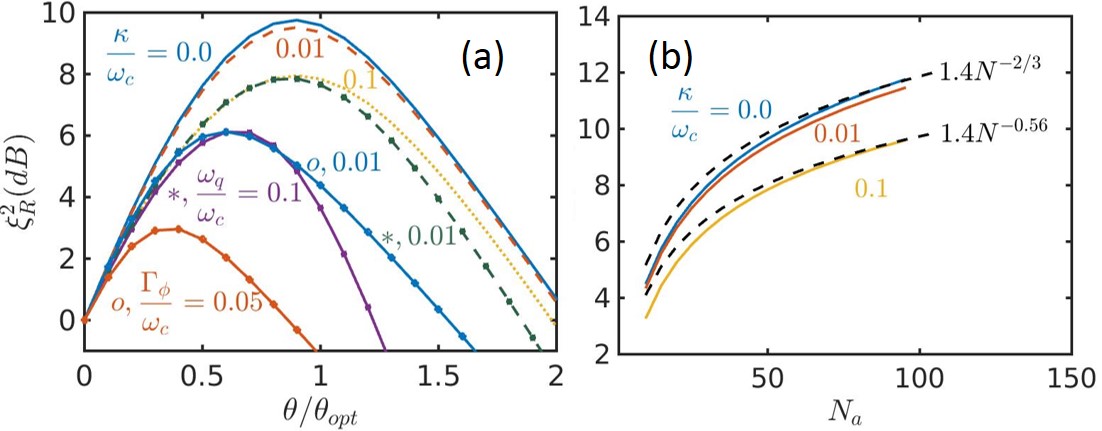} \\
  \caption{(Color online) (a) Squeezing parameter $\xi_R^2$ for $N=50$ spins as a function of the geometric phase, $\theta$, at different cavity decay rate, $\kappa$ (lines without markers, $\omega_q=\Gamma_\phi=0$), spin transition frequency, $\omega_q$ (grouped lines with $*$ markers, $\kappa=\Gamma_\phi=0$), the pure dephasing, $\Gamma_\phi$ (grouped lines with o markers, $\omega_q=\kappa=0$); (b) Squeezing parameter $\xi_R^2$ as a function of the number of spins at different $\kappa$. $\Gamma_\phi=0, \omega_q=0$ in (b).} \label{fig:xiR}
 \end{figure}

Next we go to evaluate the squeezing parameter by solving the master equation Eq. (\ref{eq:MEq}). The cavity decay and the imperfection in $\omega_q$ dominantly limit the attainable squeezing parameter. We first study the squeezing parameter for $N_a=50$ spins at time $t_1=2\pi/\omega_c$ for different ratios $\kappa/\omega_c$ and $\omega_q/\omega_c$, as shown in Fig. \ref{fig:xiR}(a). The squeezing is maximal around $\theta_\text{opt}$ in the case of small $\omega_q/\omega_c$. When $\kappa=0$ and $\omega_q=0$, we obtain $\xi_R^2=9.6 ~\deci\bel$. The squeezing parameter reduces slightly for $\kappa\leq 0.01 \omega_c$ ($\omega_q=0$). Even when the cavity decay increases to a relative large number, $\kappa= 0.1 \omega_c$, $\xi_R^2=7.7~\deci\bel$ is still achieved. In contrast, the imperfection in $\omega_q$ has stronger effect on the squeezing. The squeezing parameter for $\kappa=0$ and $\omega_q/\omega_c=0.01$ is very close to that for $\kappa/\omega_c=0.01$ and $\omega_q=0$, while it deteriorates considerably when $\omega_q$ increases to $0.1\omega_c$. In this case, the maximal available squeezing parameter decreases to $\xi_R^2=6.1~\deci\bel$ at a reduced optimal geometric phase of $\theta=0.7\theta_\text{opt}$. In experiments, we can adjust the classical driving and the detuning so that $\omega_q<0.01 \omega_c$ to guarantee the optimal squeezing at $\theta\approx \theta_\text{opt}$. 
The pure dephasing has the strongest influence on squeezing because it destroys the coherence among spins. A small pure dephasing of $\Gamma_\phi/\omega_c=0.01$ causes the maximal squeezing parameter to decrease from $\sim 10 ~\deci\bel$ at $\theta=\theta_\text{opt}$ to $6.1  ~\deci\bel$ at $\theta=0.6 \theta_\text{opt}$. When $\Gamma_\phi/\omega_c=0.05$, the maximal squeezing degree reduces by $50\%$, to $3  ~\deci\bel$.
 
 It is always desired to provide a prediction for the attainable squeezing parameter for a large ensemble. To provide such prediction, we calculate the squeezing parameter as the number of spins, see Fig. ~\ref{fig:xiR} (b). Considering $\omega_q/\omega_c\ll 1$ available in most cases, we set $\omega_q=0$ for simplicity. The squeezing parameter is well fitted by $\xi_R^2=1.4 N^{-2/3}$ when $\kappa/\omega_c\leq 0.01$. It decreases to $1.4N^{-0.56}$ with increasing the cavity decay to $\kappa/\omega_c= 0.1$. Typically, $\kappa/\omega_c\leq 0.01$ is achievable using current available experimental technology for $N_a\sim 10^6$ ultracold atoms. It means that our geometric phase control protocol can achieve a phase uncertainty $\delta\phi\propto N^{-5/6}$, approaching the Heisenberg limit of $\delta\phi\propto N^{-1}$.

In above investigation, we neglect the small decoherence terms of spins. Next, we investigate the available squeezing degree for up to $100$ spins by solving the master equation with the spin decoherence and using experimental available numbers for parameters. In doing so, we can provide a rough estimation of the achievable squeezing parameter for $10^6$ spins by fitting the numerical data. We first find the geometric phase $\theta_\text{max}$ to achieve the maximal squeezing degree for $N_a=50$ spins. It is found that $\theta_\text{max}=\theta_\text{opt}$ for cold Rb atoms, $\theta_\text{max}=0.8\theta_\text{opt}$ for BEC and $\theta_\text{max}=0.5\theta_\text{opt}$ for SiV$^-$ centers. Then we calculate the squeezing parameter as $N_a$ varying but with $\theta=\theta_\text{max}$ fixed. After simulation, we will discuss the realistic parameters for the predicted squeezing degree for each implementation. In all of three implementations, we set $\Omega_r^2/\Delta_r^2=\Omega_s^2/\Delta_s^2<0.001$ for simplicity, which are achievable as the discussion of experimental accessible parameters below. The decoherence of spins for each sample uses experimental data. 

\begin{figure}
 \centering
 \includegraphics[width=0.8\linewidth]{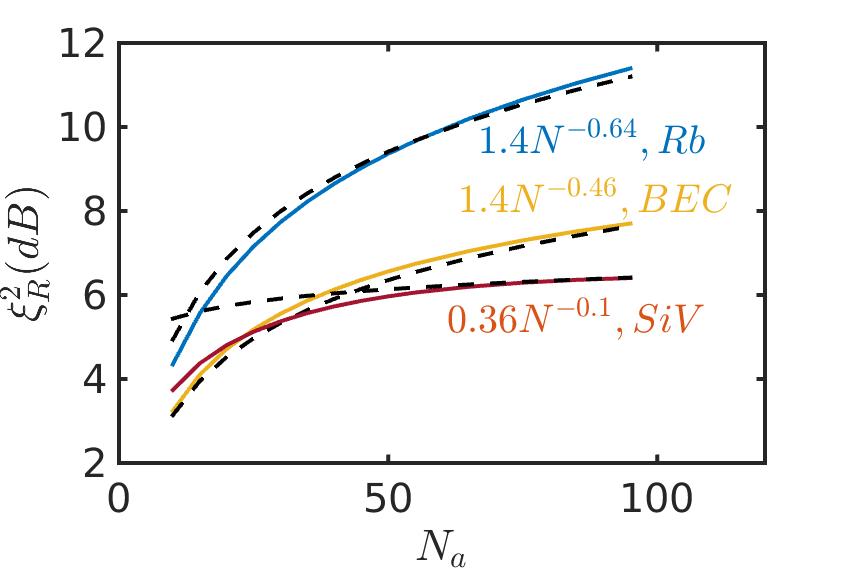} \\
 \caption{(Color online) Squeezing parameter at a particular geometric phase $\theta_\text{max}$ as a function of $N_a$ using experimental available parameters for three implementations using Rb atoms (blue line), BEC (yellow line), and SiV (Fuchsia). $\omega_c=2\pi\times 5.88 ~\mega\hertz$, $\kappa=2\pi\times 70 ~\kilo\hertz$, $\omega_q=0$, $\theta_\text{max}=\theta_\text{opt}$ for Rb atoms, $\omega_c=2\pi\times 500 ~\kilo\hertz$, $\kappa=2\pi\times 70 ~\kilo\hertz$, $\omega_q=2\pi\times 28.6~\kilo\hertz$, $\theta_\text{max}=0.8\theta_\text{opt}$ for BEC and $\omega_c=2\pi\times 350~\mega\hertz$, $\kappa=2\pi\times 1 ~\mega\hertz$, $\omega_q=0$, $\theta_\text{max}=0.5\theta_\text{opt}$ for SiV centers. The lines are fitted (black dashed lines) with $\xi_R^2= 1.4 N^{-0.64}$ for Rb atoms, $\xi_R^2= 1.4 N^{-0.46}$ for BEC and $\xi_R^2= 0.36 N^{-0.1}$ for SiV$^-$ centers.}\label{fig:ExpParam}
\end{figure}

It can be seen from see Fig. \ref{fig:ExpParam} that the largest squeezing of $\xi_R^2= 1.4 N^{-0.64}$ can be expected using an ensemble of cold alkali atoms like Rb atoms, because the total decoherence of ground states of the alkali atoms is small and the effective transition frequency $\omega_q$ can be vanishing small. Due to the large pure dephasing of SiV centers, we can only achieve squeezing of $0.36 N^{-0.1}$. According to \cite{SuperfluidGas}, the decoherence of BEC is negligible but $\omega_q=2\omega_r$ is nonzero. Taking $\omega_q=2\pi\times 28.6~\kilo\hertz$ \cite{SuperfluidGas}, we obtain the squeezing parameter of $\xi_R^2=1.4N^{-0.46}$.

Our spin-squeezing protocol via geometric phase control can be realized in various systems. For example, we can squeeze $N_a=10^6$ cold Rb atoms. 
Using the experimentally available parameters \cite{AtomicSpinSqu2,PhysRevLett.113.020408}, we choose $\omega_c=2\pi\times 5.88 ~\mega\hertz$, $\kappa=2\pi\times 70 ~\kilo\hertz$, $\omega_q=0$, $g_r=-\sqrt{3/4}g_s=2\pi \times 1.1 ~\mega\hertz$, $\Delta_s=\frac{4}{3}\Delta_r=2\pi\times 5~\giga\hertz$, $\Omega_s=\frac{\Delta_s}{50}$ and $\Omega_r=-\sqrt{\frac{3}{4}\Omega_s}$ yielding $\lambda/2\pi=-12.7~\kilo\hertz$, and $\frac{|g_s|}{\Delta_s}, \frac{|g_r|}{\Delta_r}<3\times 10^{-4}$, $\frac{\Omega_r}{2\Delta_r}\sim -1.1\times 10^{-2}, \frac{\Omega_s}{2\Delta_s}\sim -8.7\times 10^{-3}$. According to the prediction in Fig. \ref{fig:ExpParam}, the ensemble of $10^6$ Rb atoms can be squeezed by $\xi_R^2 \approx 37 ~\deci\bel$, and the phase uncertainty in measurement with squeezed spins is $\delta\phi \sim 1/N^{-0.82}$, very close to the Heisenberg limit. If we trap billion \cite{BillionAtoms1} or trillions \cite{PhysRevLett.101.073601} cold atoms in the cavity, we are potentially able to obtain a squeeze degree of $\xi_R^2=56~\deci\bel$ or even $\xi_R^2=75~\deci\bel$, respectively.
The superfluid gas has the smallest decoherence but $\omega_q=2\pi\times 28.6~\kilo\hertz$ \cite{SuperfluidGas}. We take, $\kappa=2\pi \times 70~\kilo\hertz$, $\Delta_c/2\pi=-4~\mega\hertz$, $UB/2\pi= -3.5~\mega\hertz$ yielding  $\omega_c/2\pi= 500~\kilo\hertz$, and assume $\lambda=2\pi\times 0.88~\kilo\hertz$. Correspondingly, the superfluid gas including $10^6$ ultracold atoms can be squeezed by $\xi_R^2 \approx 26 ~\deci\bel$. It is worth noting that this is the first proposal for quantum squeezing momentum of BEC. 
Our protocol can only squeeze one-million SiV$^-$ centers by $10.4 ~\deci\bel$ because SiV$^-$ centers has a pure dephasing of $\Gamma_\phi/2\pi =3.5~\mega\hertz$ \cite{PhysRevLett.113.263601,PhysRevLett.113.263602}. To achieve it, we take $\kappa=2\pi \times 1~\mega\hertz$,$\omega_c=2\pi \times 350~\mega\hertz$, $\Delta=\Delta_r=\Delta_s=2\pi\times 10~\giga\hertz$,  $\Omega_r=\Omega_s=\Delta/30$, and a large single-atom coupling $g_r=g_s=2\pi\times 46~\mega\hertz$, leading to $\frac{|g_s|}{\Delta_s}= \frac{|g_r|}{\Delta_r}=4.6\times 10^{-3}$, $\frac{\Omega_r}{2\Delta_r}=\frac{\Omega_s}{2\Delta_s}=0.017$. Such coupling strength requires a mode volume of cavity $V_c> 3000 ~\micro\meter^3$ if the dipole moment $d>10^{-29}~\coulomb \cdot \meter^3$.

Using the CARTs in spins, we have proposed a geometric phase control scheme to squeeze ensemble of spin. The available squeezing with increasing the number of spins has been numerically studied and can be tens of dB. The protocol is free of the detrimental thermal noise which heavily destroys the squeezing in mechanical resonator-based schemes. Our scheme paves a way to prepare the quantum state of a large ensemble of spins for achieving ultrasensitive quantum sensing.

The work is partly supported by the Australian Research Council Centre of Excellence for Engineered Quantum Systems (EQuS), Project No. CE110001013.
 

\begin{thebibliography}{10}

\bibitem{NatPhys.3.227}
D.~Budker and M.~Romalis.
\newblock Optical magnetometry.
\newblock {\em Nat. Phys.}, 3:227--234, 2007.

\bibitem{NatPhys.10.21}
Fazhan Shi, Xi~Kong, Pengfei Wang, Fei Kong, Nan Zhao, Ren-Bao Liu, and
  Jiangfeng Du.
\newblock Sensing and atomic-scale structure analysis of single nuclear-spin
  clusters in diamond.
\newblock {\em Nat. Phys.}, 10:21--25, 2014.

\bibitem{Science.339.561}
T.~Staudacher, F.~Shi, S.~Pezzagna, J.~Meijer, J.~Du, C.~A. Meriles,
  F.~Reinhard, and J.~Wrachtrup.
\newblock Nuclear magnetic resonance spectroscopy on a
  $($5$-\text{Nanometer})^3$ sample volume.
\newblock {\em Science}, 339:561--563, 2013.

\bibitem{PhysRevX.4.021045}
R.~J. Sewell, M.~Napolitano, N.~Behbood, G.~Colangelo, F.~Martin~Ciurana, and
  M.~W. Mitchell.
\newblock Ultrasensitive atomic spin measurements with a nonlinear
  interferometer.
\newblock {\em Phys. Rev. X}, 4:021045, 2014.

\bibitem{PhysRevLett.110.160802}
D.~Sheng, S.~Li, N.~Dural, and M.~V. Romalis.
\newblock Subfemtotesla scalar atomic magnetometry using multipass cells.
\newblock {\em Phys. Rev. Lett.}, 110:160802, 2013.

\bibitem{PhysRevLett.112.160802}
K.~Jensen, N.~Leefer, A.~Jarmola, Y.~Dumeige, V.~M. Acosta, P.~Kehayias,
  B.~Patton, and D.~Budker.
\newblock Cavity-enhanced room-temperature magnetometry using absorption by
  nitrogen-vacancy centers in diamond.
\newblock {\em Phys. Rev. Lett.}, 112:160802, 2014.

\bibitem{PhysRevLett.110.130802}
Kejie Fang, Victor~M. Acosta, Charles Santori, Zhihong Huang, Kohei~M. Itoh,
  Hideyuki Watanabe, Shinichi Shikata, and Raymond~G. Beausoleil.
\newblock High-sensitivity magnetometry based on quantum beats in diamond
  nitrogen-vacancy centers.
\newblock {\em Phys. Rev. Lett.}, 110:130802, 2013.

\bibitem{PhysRevX.5.041001}
Thomas Wolf, Philipp Neumann, Kazuo Nakamura, Hitoshi Sumiya, Takeshi Ohshima,
  Junichi Isoya, and J\"org Wrachtrup.
\newblock Subpicotesla diamond magnetometry.
\newblock {\em Phys. Rev. X}, 5:041001, 2015.

\bibitem{NatCommun.6.8251}
Liang Jin, Matthias Pfender, Nabeel Aslam, Philipp Neumann, Sen Yang, J\"{o}rg
  Wrachtrup, and Ren-Bao Liu.
\newblock Proposal for a room-temperature diamond maser.
\newblock {\em Nat. Commun.}, 6:8251, 2015.

\bibitem{PhysRevA.92.043409}
Keyu Xia, Nan Zhao, and Jason Twamley.
\newblock Detection of a weak magnetic field via cavity-enhanced faraday
  rotation.
\newblock {\em Phys. Rev. A}, 92:043409, 2015.

\bibitem{Science.306.1330}
Vittorio Giovannetti, Seth Lloyd, and Lorenzo Maccone.
\newblock Quantum-enhanced measurements: Beating the standard quantum limit.
\newblock {\em Science}, 306:1330, 2004.

\bibitem{Science.344.1486}
Sydney Schreppler, Nicolas Spethmann, Nathan Brahms, Thierry Botter, Maryrose
  Barrios, and Dan~M. Stamper-Kurn.
\newblock Optically measuring force near the standard quantum limit.
\newblock {\em Science}, 344:1486--1489, 2014.

\bibitem{PhysRep.509.89}
Jian Ma, Xiaoguang Wang, C.~P. Sun, and Franco Nori.
\newblock Quantum spin squeezing.
\newblock {\em Phys. Rep.}, 509:89--165, 2011.

\bibitem{AtomicSpinSqu2}
Onur Hosten, Nils~J. Engelsen, Rajiv Krishnakumar, and Mark~A. Kasevich.
\newblock Measurement noise $100$ times lower than the quantum-projection limit
  using entangled atoms.
\newblock {\em Nature}, 529:505, 2016.

\bibitem{AtomicSpinSqu1}
C.~D. Hamley, C.~S. Gerving, T.~M. Hoang, E.~M. Bookjans, and M.~S. Chapman.
\newblock Spin-nematic squeezed vacuum in a quantum gas.
\newblock {\em Nat. Phys.}, 8:305, 2012.

\bibitem{AtomicSpinSqu4}
T.~Fernholz, H.~Krauter, K.~Jensen, J.~F. Sherson, A.~S. S\o{}rensen, and E.~S.
  Polzik.
\newblock Spin squeezing of atomic ensembles via nuclear-electronic spin
  entanglement.
\newblock {\em Phys. Rev. Lett.}, 101:073601, 2008.

\bibitem{AtomicSpinSqu3}
Emanuele~G. Dalla~Torre, Johannes Otterbach, Eugene Demler, Vladan Vuletic, and
  Mikhail~D. Lukin.
\newblock Dissipative preparation of spin squeezed atomic ensembles in a steady
  state.
\newblock {\em Phys. Rev. Lett.}, 110:120402, 2013.

\bibitem{PhysRevLett.109.173603}
Leigh~M. Norris, Collin~M. Trail, Poul~S. Jessen, and Ivan~H. Deutsch.
\newblock Enhanced squeezing of a collective spin via control of its qudit
  subsystems.
\newblock {\em Phys. Rev. Lett.}, 109:173603, 2012.

\bibitem{CavityFeedback1}
Ian~D. Leroux, Monika~H. Schleier-Smith, and Vladan
  Vuleti\ifmmode~\acute{c}\else \'{c}\fi{}.
\newblock Orientation-dependent entanglement lifetime in a squeezed atomic
  clock.
\newblock {\em Phys. Rev. Lett.}, 104:250801, 2010.

\bibitem{CavityFeedback2}
Ian~D. Leroux, Monika~H. Schleier-Smith, and Vladan
  Vuleti\ifmmode~\acute{c}\else \'{c}\fi{}.
\newblock Implementation of cavity squeezing of a collective atomic spin.
\newblock {\em Phys. Rev. Lett.}, 104:073602, 2010.

\bibitem{PhysRevLett.110.156402}
S.~D. Bennett, N.~Y. Yao, J.~Otterbach, P.~Zoller, P.~Rabl, and M.~D. Lukin.
\newblock Phonon-induced spin-spin interactions in diamond nanostructures:
  Application to spin squeezing.
\newblock {\em Phys. Rev. Lett.}, 110:156402, 2013.

\bibitem{PhysRevA.92.013825}
Yan-Lei Zhang, Chang-Ling Zou, Xu-Bo Zou, Liang Jiang, and Guang-Can Guo.
\newblock Phonon-induced spin squeezing based on geometric phase.
\newblock {\em Phys. Rev. A}, 92:013825, 2015.

\bibitem{KXJT}
Keyu Xia and Jason Twamley.
\newblock {Generating spin squeezing states and GHZ entanglement using a hybrid
  phonon-spin ensemble in diamond}.
\newblock 2016.

\bibitem{PhysRevA.75.013804}
F.~Dimer, B.~Estienne, A.~S. Parkins, and H.~J. Carmichael.
\newblock Proposed realization of the dicke-model quantum phase transition in
  an optical cavity qed system.
\newblock {\em Phys. Rev. A}, 75:013804, 2007.

\bibitem{PhysRevLett.113.020408}
Markus~P. Baden, Kyle~J. Arnold, Arne~L. Grimsmo, Scott Parkins, and Murray~D.
  Barrett.
\newblock Realization of the dicke model using cavity-assisted raman
  transitions.
\newblock {\em Phys. Rev. Lett.}, 113:020408, 2014.

\bibitem{SuperfluidGas}
Kristain Baumann, Christine Guerlin, Ferdinand Brennecke, and Tilman Esslinger.
\newblock Dicke quantum phase transition with a superfluid gas in an optical
  cavity.
\newblock {\em Nature}, 464:1301--1306, 2010.

\bibitem{AtomData}
D.~Steck.
\newblock http://steck.us/alkalidata.
\newblock revision 2.1.5, January 13, 2015.

\bibitem{PhysRevLett.113.263601}
Benjamin Pingault, Jonas~N. Becker, Carsten H.~H. Schulte, Carsten Arend,
  Christian Hepp, Tillmann Godde, Alexander~I. Tartakovskii, Matthew Markham,
  Christoph Becher, and Mete Atat\"ure.
\newblock All-optical formation of coherent dark states of silicon-vacancy
  spins in diamond.
\newblock {\em Phys. Rev. Lett.}, 113:263601, 2014.

\bibitem{PhysRevLett.113.263602}
Lachlan~J. Rogers, Kay~D. Jahnke, Mathias~H. Metsch, Alp Sipahigil, Jan~M.
  Binder, Tokuyuki Teraji, Hitoshi Sumiya, Junichi Isoya, Mikhail~D. Lukin,
  Philip Hemmer, and Fedor Jelezko.
\newblock All-optical initialization, readout, and coherent preparation of
  single silicon-vacancy spins in diamond.
\newblock {\em Phys. Rev. Lett.}, 113:263602, 2014.

\bibitem{PhysRevLett.112.036405}
Christian Hepp, Tina M\"uller, Victor Waselowski, Jonas~N. Becker, Benjamin
  Pingault, Hadwig Sternschulte, Doris Steinm\"uller-Nethl, Adam Gali,
  Jeronimo~R. Maze, Mete Atat\"ure, and Christoph Becher.
\newblock Electronic structure of the silicon vacancy color center in diamond.
\newblock {\em Phys. Rev. Lett.}, 112:036405, 2014.

\bibitem{NJP.17.043011}
Kay~D Jahnke, Alp Sipahigil, Jan~M Binder, Marcus~W Doherty, Mathias Metsch,
  Lachlan~J. Rogers, Neil~B. Manson, Mikhail~D. Lukin, and Fedor Jelezko.
\newblock Electron-phonon processes of the silicon-vacancy centre in diamond.
\newblock {\em New J. Phys.}, 17:043011, 2015.

\bibitem{SupplementaryInf}
See the supplementary information.

\bibitem{HPTransf}
T.~Holstein and H.~Primakoff.
\newblock Field dependence of the intrinsic domain magnetization of a
  ferromagnet.
\newblock {\em Phys. Rev.}, 58:1098--1113, 1940.

\bibitem{DickeQPT1}
D.~Nagy, G.~K\'onya, G.~Szirmai, and P.~Domokos.
\newblock Dicke-model phase transition in the quantum motion of a bose-einstein
  condensate in an optical cavity.
\newblock {\em Phys. Rev. Lett.}, 104:130401, 2010.

\bibitem{MagnusFormula}
W.~Magnus.
\newblock On the exponential solution of differential equations for a linear
  operator.
\newblock {\em Commun. Pure Appl. Math.}, 7:649, 1954.

\bibitem{PhysRevLett.90.160402}
A.~Carollo, I.~Fuentes-Guridi, M.~Fran\ifmmode \mbox{\c{c}}\else~\c{c}\fi{}a
  Santos, and V.~Vedral.
\newblock Geometric phase in open systems.
\newblock {\em Phys. Rev. Lett.}, 90:160402, 2003.

\bibitem{RamanModel}
Florentin Reiter and Anders~S. S\o{}rensen.
\newblock Effective operator formalism for open quantum systems.
\newblock {\em Phys. Rev. A}, 85:032111, 2012.

\bibitem{BillionAtoms1}
A.~Ridinger, S.~Chaudhuri, T.~Salez, U.~Eismann, D.~R. Fernandes, K.~Magalh\
  {a}es, D.~Wilkowski, C.~Salomon, and F.~Chevy.
\newblock Large atom number dual-species magneto-optical trap for fermionic
  $^6$li and $^40$k atoms.
\newblock {\em Eur. Phys. J. D}, 65:223, 2011.

\bibitem{PhysRevLett.101.073601}
T.~Fernholz, H.~Krauter, K.~Jensen, J.~F. Sherson, A.~S. S\o{}rensen, and E.~S.
  Polzik.
\newblock Spin squeezing of atomic ensembles via nuclear-electronic spin
  entanglement.
\newblock {\em Phys. Rev. Lett.}, 101:073601, 2008.

\end{thebibliography}


\end{document}